# A Divide-and-Conquer Machine Learning Approach for Modelling Turbulent Flows

Anthony Man[1], Mohammad Jadidi[1], Amir Keshmiri[1], Hujun Yin[2], Yasser Mahmoudi[1*]

[1] Department of Mechanical, Aerospace and Civil Engineering (MACE), The University of Manchester, M13 9PL, UK
[2] Department of Electrical and Electronic Engineering, The University of Manchester, M13 9PL, UK

*Corresponding author: yasser.mahmoudilarimi@manchester.ac.uk

**Abstract**

In this paper, a novel zonal machine learning (ML) approach for Reynolds-averaged Navier-Stokes (RANS) turbulence modelling based on the divide-and-conquer technique is introduced. This approach involves partitioning the flow domain into regions of flow physics called zones, training one ML model in each zone, then validating and testing them on their respective zones. The approach was demonstrated with the tensor basis neural network (TBNN) and another neural net called the turbulent kinetic energy neural network (TKENN). These were used to predict Reynolds stress anisotropy and turbulent kinetic energy respectively in test cases of flow over a solid block, which contain regions of different flow physics including separated flows. The results show that the combined predictions given by the zonal TBNNs and TKENNs were significantly more accurate than their corresponding standard non-zonal models. Most notably, shear anisotropy component in the test cases was predicted at least 20% and 55% more accurately on average by the zonal TBNNs compared to the non-zonal TBNN and RANS, respectively. The Reynolds stress constructed with the zonal predictions was also found to be at least 23% more accurate than those obtained with the non-zonal approach and 30% more accurate than the Reynolds stress predicted by RANS on average. These improvements were attributed to the shape of the zones enabling the zonal models to become highly locally optimized at predicting the output.

# Nomenclature

| Variable | Meaning | Unit |
|---|---|---|
| $a_{ij}$ | Anisotropic component of Reynolds stress | m²/s² |
| $b_{ij}$ | Anisotropy tensor | - |
| $g^{(n)}$ | Scalar coefficients of general effective viscosity hypothesis | - |
| h | Step height | m |
| $k$ | Turbulent kinetic energy | m²/s² |
| $N_s$ | Number of data samples | - |
| Re | Reynolds number (based on bulk inlet velocity $U_b$ and block height h) | - |
| $R_{ij}$ | Mean rotation rate | 1/s |
| $S_{ij}$ | Mean strain rate | 1/s |
| $T^{(n)}$ | Tensor invariants of general effective viscosity hypothesis | - |
| $U_b$ | Bulk inlet velocity | m/s |
| $u'_i$ | Velocity fluctuation in i$^{th}$ direction | m/s |
| x | Streamwise direction | m |
| y | Vertical (wall-normal) direction | m |
| Y | Turbulent kinetic energy in logarithm form ($Y = \log_{10}(k)$) | - |

| Symbol | | |
|---|---|---|
| $\alpha$ | Non-dimensional turbulent kinetic energy ($\alpha = k/U_b^2$) | - |
| $\tau_{ij}$ | Reynolds stress | m²/s² |
| $\delta_{ij}$ | Kronecker delta | |
| $\nu_t$ | Kinematic eddy viscosity | m²/s |
| $\varepsilon$ | Turbulent kinetic energy dissipation rate | m²/s³ |
| $\gamma$ | Learning rate scheduler factor | - |
| ω | Specific turbulent kinetic energy dissipation rate | 1/s |

**Subscript**
*Pred*     Predicted
*x*     Streamwise direction

**Superscript**
$^*$     Non-dimensionalized with $k/\varepsilon$
$\overline{\square}$     Reynolds-averaged

**Abbreviation**
BC     Boundary condition
BM     Barycentric map
DaC     Divide-and-conquer
DNS     Direct numerical simulation
ELU     Exponential linear unit
GEVH     General effective-viscosity hypothesis
LES     Large eddy simulation
LEVM     Linear eddy viscosity model
LR     Learning rate
ML     Machine learning
MoE     Mixture of experts

| | |
|---|---|
| MSE | Mean squared error |
| NN | Neural network |
| RAI | RANS accuracy improvement |
| RANS | Reynolds-averaged Navier-Stokes |
| ReLU | Rectified linear unit |
| SGS | Subgrid scale |
| SST | Shear stress transport |
| TBNN | Tensor basis neural network |
| TKE | Turbulent kinetic energy |
| TKENN | Turbulent kinetic energy neural network |

# 1 Introduction

For decades, intensive research effort has been dedicated towards more accurate, low computational cost turbulence modelling. However, due to a current plateau in progress, turbulent flows are still often modelled using Reynolds-averaged Navier-Stokes (RANS) approaches that give unsatisfactory predictions for complex flow phenomena [1]. This is largely caused by inaccurate modelling of the Reynolds stresses [2]. Popular RANS models such as k-ω shear stress transport (SST) employ a linear relationship between mean strain rate and Reynolds stress, hence these are known as linear eddy viscosity models (LEVMs) [3]. This linear relationship along with continuity enforces constraints on the relative magnitudes of the normal Reynolds stresses, and turbulent kinetic energy (TKE) which represents an aggregation of these is normally used instead to solve the RANS equations. As a result, LEVMs give satisfactory predictions for flows where the gradients of normal Reynolds stresses are negligible or isotropic [3]. However, the opposite occurs in some complex turbulent flow phenomena such as separated flows, which are present in many engineering problems, including boundary layer detachment from airfoil surfaces, and airflow over vehicles, buildings, and terrain.

Non-linear eddy viscosity models have been proposed in the past which sensitize the Reynolds stress to both mean strain and rotation rate, as well as their higher order products [2, 4, 5]. This allows the normal Reynolds stresses to be unconstrained, and thus more accurate predictions of mean velocity and reattachment length by these models have been reported in the literature. Nonetheless, the empirical tuning of numerous coefficients in these models can be a lengthy process and can result in the model being sensitive to certain flow types [6]. Reynolds stress transport models are an alternative, but they currently do not consistently predict more accurately than LEVMs and lack robustness due to their complexity [7]. Lastly, given that the "age" of Moore's law is coming to an end, it is reasonable to suggest that high fidelity methods such as large eddy simulation (LES) and direct numerical simulation (DNS) will not be computationally feasible for most CFD practitioners to use in the near future [1, 8].

More recently, advancement in machine learning (ML) has propelled the use of data-driven methods for modelling Reynolds stress [9, 10]. Many studies on these have reported promising results – with accuracy of predictions closer to that of LES or DNS but still retaining computational cost on the same order of magnitude as RANS simulations. As many turbulent flows are still being simulated with LEVMs, we focus the following literature review on existing ML approaches that take quantities from LEVM simulations as model inputs to predict an improved Reynolds stress (or its projections). Pope [11] showed that the Reynolds stress in a homogeneous turbulent flow can be expressed as a tensor integrity basis, and its unknown scalar coefficients are dependent on the mean strain and rotation rate invariants. Ling et al. [12] trained a neural network (NN) to predict these scalar coefficients, which gave promising Reynolds stress anisotropy and mean flow field results when tested on canonical flow problems. This original approach fully accounts for both mean strain and rotation rate, as well as their higher order terms in the calculation of Reynolds stress anisotropy. They called their NN the Tensor basis neural network (TBNN), and their work inspired many subsequent studies [13-17]. Different ML techniques have been used to predict the scalars, including symbolic regression [18-20] and random forests [21]. Some studies have chosen to predict different forms of the Reynolds stress instead, such as its Eigendecomposition components [22-26], divergence [27] or linear and nonlinear parts [28].

Many of the aforementioned investigations used flow over periodic hills as training and test cases for their ML models. These cases contain multiple flow physics situated in isolated regions of the domain, namely separated flow and turbulent boundary layers. Studies that used these cases for testing have reported prediction discrepancies in one of the regions, despite an excellent match with "true" results in the other [16, 22, 24, 28]. Hence, one ML model inherently cannot be fitted well to predict both regions accurately. This is an especially concerning issue for problems containing separated flows, as many of these have other flow physics, *e.g*., von Karman vortex street in flow past a cylinder. Predicting all regions of physics accurately can ensure that the separated flow region is uncompromised. Yet this issue is not isolated to separated flow problems, as many contain more than one type of flow physics. To progress in this direction of improving generalizability in one brute-force model, the following bottlenecks must be addressed: (i) increasing the amount of training data to encompass different turbulent regimes and flow behaviours for (ii) training a ML model with complex architecture that may accurately account for many degrees of freedom in these various flows [29, 30]. Given current computational capabilities, addressing both will be challenging to achieve in the near future.

Although producing accurate ML predictions for cases with multiple flow physics is a complex task as discussed, it is naturally suited to be divided into subtasks, as each type of flow physics usually only occurs in a specific region. The domain may be decomposed into these regions and one ML model per region may be trained, validated, and tested on data from them. In theory, this partially negates the two bottlenecks, as less training data can be used to converge each ML model and the complexity of the models can be reduced when only accounting for one type of flow physics [31]. Additionally, each model can become optimized at predicting its respective region, thus leading to the entire domain being well predicted. This is a divide-and-conquer (DaC) approach – an algorithm paradigm which should be familiar to most ML or CFD practitioners as it is applied in both fields, *e.g*., in mixture of experts and parallel CFD computation [32]. Utilizing separate ML models in different regions for a simple turbulent flow has been attempted by Zhu et al. [33, 34], who reconstructed a mapping function between the turbulent eddy viscosity and the mean flow variables for RANS turbulence models. However, their studies were performed for simple canonical turbulent flow over airfoils with no flow separation. Additionally, they used RANS results as the ground-truth data for training the ML algorithm, and consequently their ML models could not improve the accuracy of RANS prediction for turbulent flows.

The above literature review shows that the application of ML for RANS closure modelling has gained great traction over recent years, due to advancements in ML and a plateau in accuracy improvements from traditional RANS development. However, for problems involving multiple regions with complex flow physics, it is now demonstrated that the conventional ML models cannot predict more than one region accurately. This is because the models have been trained with an insufficient amount of data or are not sophisticated enough to accurately account for all the different flow physics. Addressing these requirements with current computational capabilities is challenging. To partially negate them, a novel zonal ML approach based on the divide-and-conquer technique is proposed for improving RANS predictions – consisting of using multiple ML models, with each trained, validated, and tested on data from a specific region of physics in the flow domain. The approach is demonstrated using data from RANS and well-resolved LES cases of flow over a solid block, which contain several distinct regions of flow physics including separated flows. The main objective of this paper is to show that the zonal approach is able to predict all components of Reynolds stress anisotropy, TKE and Reynolds stress more accurately than when one ML model is deployed on the whole flow domain, such as in the TBNN.

## 2 Methodology

When performing a RANS simulation with a LEVM, Reynolds stress $\tau_{ij}$ is traditionally modelled using Boussinesq Hypothesis (BH) to close the system of RANS equations. For incompressible flows, BH is expressed as [35]:

$$\tau_{ij} = \overline{u'_i u'_j} = \frac{2}{3}k\delta_{ij} - \underbrace{2\nu_t S_{ij}}_{a_{ij}} \quad (1)$$

where, $k$ is the turbulent kinetic energy (TKE), $\delta_{ij}$ is Kronecker delta, $\nu_t$ is the kinematic eddy viscosity and $S_{ij}$ is the mean strain rate tensor. The anisotropic component of $\tau_{ij}$ denoted $a_{ij}$ in Eq. (1) is directly responsible for turbulent transport, whereas the isotropic $2k\delta_{ij}/3$ part is usually combined with the pressure term in the RANS equations. $a_{ij}$ can be non-dimensionalized to give [11]:

$$b_{ij} \equiv \frac{\tau_{ij}}{2k} - \frac{1}{3}\delta_{ij} \quad (2)$$

where, $b_{ij}$ will be referred to as the anisotropy tensor hereafter. Comparing Eq. (1) with Eq. (2) shows that BH models $b_{ij}$ as $-\nu_t S_{ij}/k$. Pope [11] showed that this is a gross simplification that yields inaccurate predictions in many flows, including those with high strain and rotation effects. To account for these on $b_{ij}$, Pope [11] postulated the general effective-viscosity hypothesis (GEVH), which is a complete finite tensor polynomial expression for $b_{ij}$:

$$b_{ij} = \sum_{n=1}^{10} g^{(n)} T^{(n)} \qquad (3)$$

$g^{(n)}$ are scalars that are unknown functions of the non-dimensional mean strain $S_{ij}^*(= kS_{ij}/\varepsilon)$ and rotation rate $R_{ij}^*(= kR_{ij}/\varepsilon)$ invariants, where $\varepsilon$ is the TKE dissipation rate. $T^{(n)}$ are known tensor functions of $S_{ij}^*$ and $R_{ij}^*$. Readers are referred to Appendix A for more information.

## 2A  Neural Networks (NNs)

Ling et al. [12] showed that Eq. (3) can be modelled as a NN to predict $b_{ij}$ locally. Their original model called the Tensor Basis Neural Network (TBNN) is shown in **Figure 1**. $S_{ij}^*$ and $R_{ij}^*$ invariants calculated from RANS data for a single location are fed into the invariant input layer, and these values pass through feedforward fully connected hidden layers to produce predictions of $g^{(n)}$ at the final hidden layer. These are multiplied elementwise with the tensor input layer, which consists of $T^{(n)}$ tensors calculated using RANS data from the same location to give predictions of $b_{ij}$ at the merge output layer. While there are other ML methods for predicting $b_{ij}$ and other choices of Reynolds stress projections that can be predicted, we chose to demonstrate the zonal approach with TBNN as it is based on mathematical foundations in GEVH and has been widely used. In addition, $b_{ij}$ has the following advantages over other Reynolds stress projections as an ML output: (i) it is Reynolds number invariant, *i.e.* at different Reynolds numbers, the same $b_{ij}$ field can be obtained for the same fluid flow system and geometry, and this therefore provides good Reynolds number extrapolation capability in ML models [17] and (ii) its dimensional form $a_{ij}$ is often used in RANS solvers rather than $\tau_{ij}$ itself. Therefore $b_{ij}$ predictions from ML models can be easily inserted back into CFD cases for a posteriori investigation. The standard inputs of $S_{ij}^*$ and $R_{ij}^*$ invariants shown in Appendix A were used in this work.

Anisotropy tensor $b_{ij}$ outputs from TBNNs can be converted into $\tau_{ij}$ predictions by substituting them into Eq. (2) and rearranging for $\tau_{ij}$. This shows that $k$ must be given to complete the conversion to $\tau_{ij}$ predictions. As $\tau_{ij}$ from LES was used to calculate "true" $b_{ij}$, the values of $k$ must also be close to those predicted by LES to give $\tau_{ij}$ of improved accuracy. As "true" $k$ results are not accessible during TBNN deployment, $k$ must therefore be predicted with a separate ML model. The input variables of this model must only consist of quantities obtainable from RANS so that it can be deployed in the absence of LES, DNS, or experimental results. Some studies in the literature have reported on training a ML model to predict $k$ for this purpose [24, 25]. Aside from invariants of $S_{ij}^*$ and $R_{ij}^*$, these studies also included scalar markers and other invariants in their input features. In this work, the $S_{ij}^*$ and $R_{ij}^*$ invariants used as TBNN inputs were also applied in this ML model for consistency. In addition, $\log_{10}(k_{RANS})$ was used as a supplementary input feature so that the model could be sensitized to a baseline TKE distribution, where $k_{RANS}$ is the TKE modelled by RANS. The log scaling normalization on $k_{RANS}$ reduces its distribution skewness, which helps to improve model performance, while transforming $k_{RANS}$ to give a similar magnitude range as the $S_{ij}^*$ and $R_{ij}^*$ invariants – thus allowing all input features to be on a similar scale so that no feature would dominate model training [36]. To ensure that predicted $k$ values were non-negative, we follow the principle by Wu et al. [24] and assign $\log_{10}(k)$ as the output. For consistency with the TBNN, a feedforward fully connected NN was chosen for the ML model type as shown in **Figure 1**, which hereafter will be referred to as the TKE neural network (TKENN). Likewise with TBNN, the TKENN was also used to demonstrate the zonal approach.

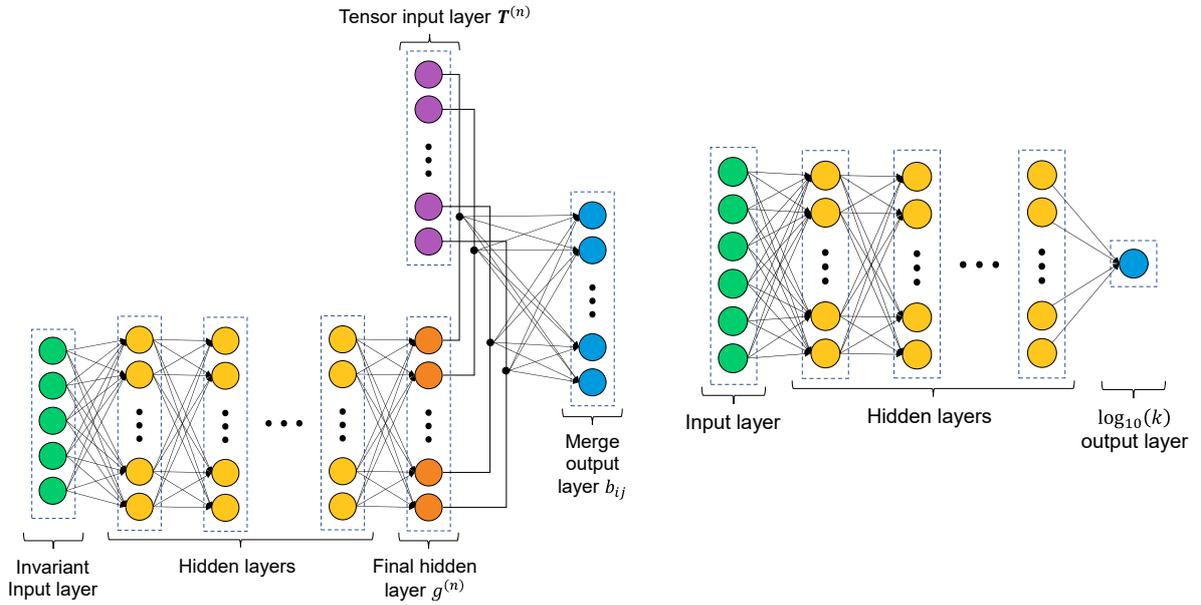

**Figure 1 left:** TBNN and **right:** TKENN architecture [12].

## 2B  Flow Separation Over a Solid Block

Data from cases of flow over a solid block in a channel was used to assess the zonal approach. These cases contain several distinct regions of flow physics labelled in **Figure 2** as follows: (1) laminar flow upstream of the block, (2) a developing turbulent boundary layer on the upper wall, (3) separated flow over the block leading edge, creating vortical structures (**Figure 3**) and a recirculation region, (4) boundary layer redevelopment above the block, (5) flow separation from the block trailing edge, creating a wake with reversed flow, and (6) boundary layer redevelopment downstream of the wake [37]. The results of these cases can therefore offer clear zonal partitions. Having discussed the current unfeasibility of high-fidelity CFD methods and the shortcomings of RANS when predicting separated flows, one may therefore choose to use ML in such flow problems. For this reason, we intentionally demonstrate the zonal approach on these cases where flow separations dominate the fluid dynamics.

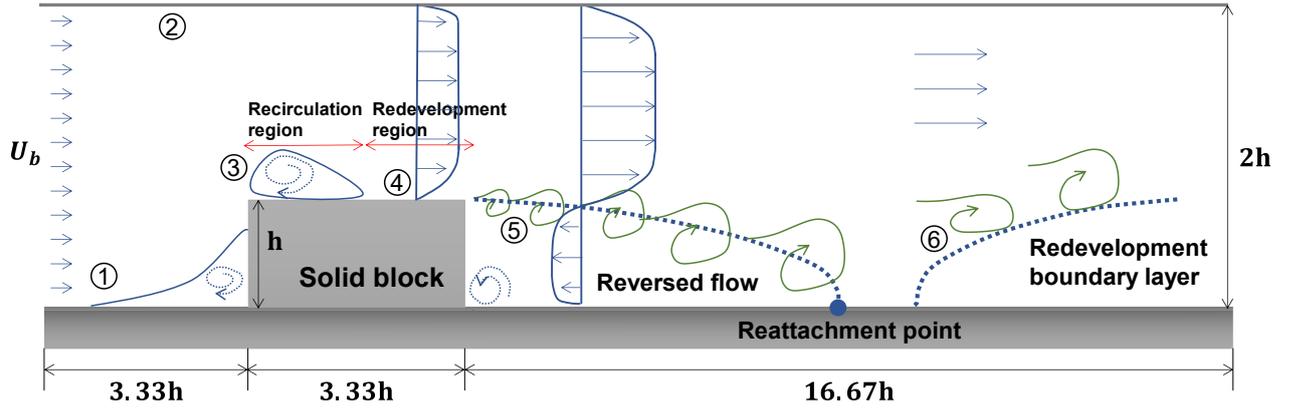

**Figure 2** A schematic diagram of flow over a solid block in a channel.

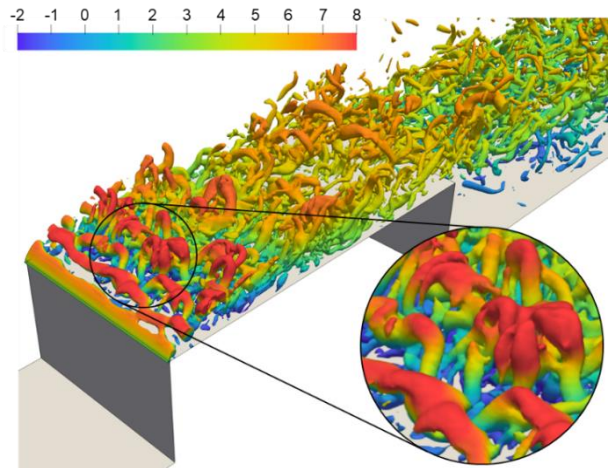

**Figure 3** Iso-volume plot of Q-criterion = $2\times10^6$ s$^{-2}$ colored by mean streamwise velocity $U_x$ (m/s) for flow over a solid block at Re = 5400 ($U_b$ = 3 m/s) simulated using LES. This plot shows the hairpin vortical structures developed downstream of the block leading edge.

To illustrate the shortcomings of RANS approaches when simulating these flow cases, contours of TKE predicted by RANS and LES for Re = 5400 (based on bulk inlet velocity $U_b$ = 3 m/s and block height h = 18 mm) are shown in **Figure 4**. The RANS simulation was performed with the k-ω SST model as it remains widely used in industry. A typical TKE result modelled by RANS for this problem is therefore presented. **Figure 4** shows that RANS severely underpredicts TKE, especially in the high TKE regions caused by separated flow. In addition, the location of maximum TKE above the block predicted by RANS is delayed by approximately 0.5h compared to the LES result. This delayed maxima, low TKE leading to underpredicted eddy viscosity, and weak TKE gradients all contribute towards over-predicted reattachment lengths – in this case by 83% above the block and 40% in the downstream wake as shown in **Figure 5** [38-40]. These results show that TKE is highly responsible for the inaccurate modelling of velocity in separated flows when using RANS approaches. Due to the anisotropic nature of the normal Reynolds stresses in flow separations, it is not appropriate to assume they are isotropic and to use TKE as their aggregate form to solve the RANS equations in these cases [41]. This justifies our aim of developing a zonal approach for improving the accuracy of RANS-predicted Reynolds stresses, and these will be injected back into the RANS solver to assess velocity in future a posteriori work.

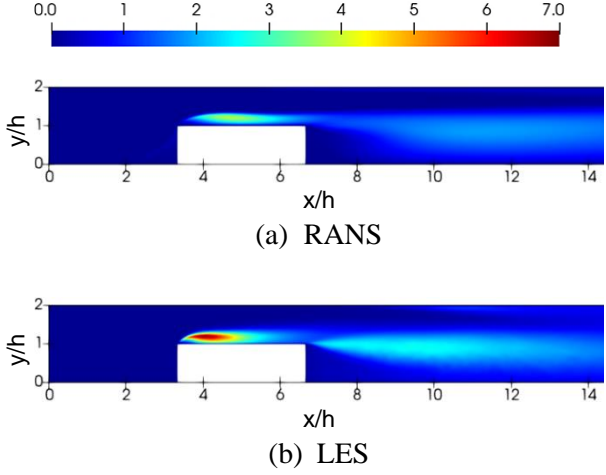

**Figure 4** Contour plots of TKE, $k$ (m$^2$/s$^2$) for flow over a solid block at Re = 5400 simulated using **(a)** RANS and **(b)** LES.

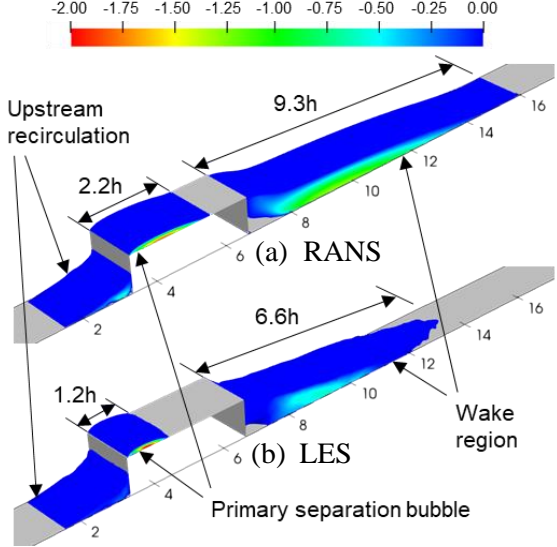

**Figure 5** Iso-volume plots showing negative mean streamwise velocity $U_x$ (m/s) for flow over a solid block at Re = 5400 simulated using **(a)** RANS and **(b)** LES.

## 2C  Simulations and Data Pre-processing

RANS simulations and well-resolved LES of flow over a solid block for the following five bulk inlet velocities were performed to produce ML datasets: 1, 2, 2.5, 3 and 4 m/s, with corresponding Reynolds numbers (based on a block height of 18 mm) of Re = [1800, 3600, 4500, 5400 and 7200]. The computational domain was three dimensional, with dimensions given in **Figure 2** and a spanwise width of 1.67h. After performing a grid independence study, all RANS simulations were run with a structured mesh consisting of 4 million elements. Uniform velocity and zero gradient pressure were used as the inlet boundary conditions (BCs). At the outlet, velocity was set to zero gradient and a fixed value of uniform zero pressure was used. The no-slip condition was applied at the walls and symmetry BCs were used in the spanwise direction. The k-ω SST turbulence model was chosen [42] with the SIMPLEC algorithm in OpenFOAM to run these steady state simulations [3, 43]. The LES cases were run with a finer structured mesh consisting of 10.5 million elements. Sub-grid scale (SGS) eddy viscosity was modelled with the dynamic SGS TKE model [44] and the same BCs as the RANS simulations were used. The PISO algorithm in OpenFOAM was chosen to converge the pressure and velocity fields [3, 43]. More information on the LES, including validation of the solver can be found in Jadidi et al. [37].

Data from the CFD cases was extracted to be read by the ML code. In particular, $k$, $\varepsilon$, velocity gradients $\nabla U$ and $\tau_{ij}$ were extracted from the spanwise centre-plane of RANS results. The first three variables were used to construct the input features, whilst the $\tau_{ij}$ results were used to calculate $b_{ij}$ from RANS to act as a baseline for TBNN results comparison. Only $\tau_{ij}$ was extracted from the centre-plane of time-averaged LES results in order for ground-truth values to be calculated for the NNs. Ground-truth TKE was calculated as $k_{LES} = 0.5 \times (\tau_{11,LES} + \tau_{22,LES} + \tau_{33,LES})$, and Eq. (2) was used with $k_{LES}$ and $\tau_{ij,LES}$ to calculate ground-truth $b_{ij}$ values for the TBNNs, whilst the logarithm operation was directly applied on $k_{LES}$ to give output target values for the TKENNs. The RANS grid centre-plane contained $8 \times 10^4$ cell-centre locations, hence there were $8 \times 10^4$ rows of input and output features for each Reynolds number. **Figure 6** demonstrates all stages of the present ML workflow.

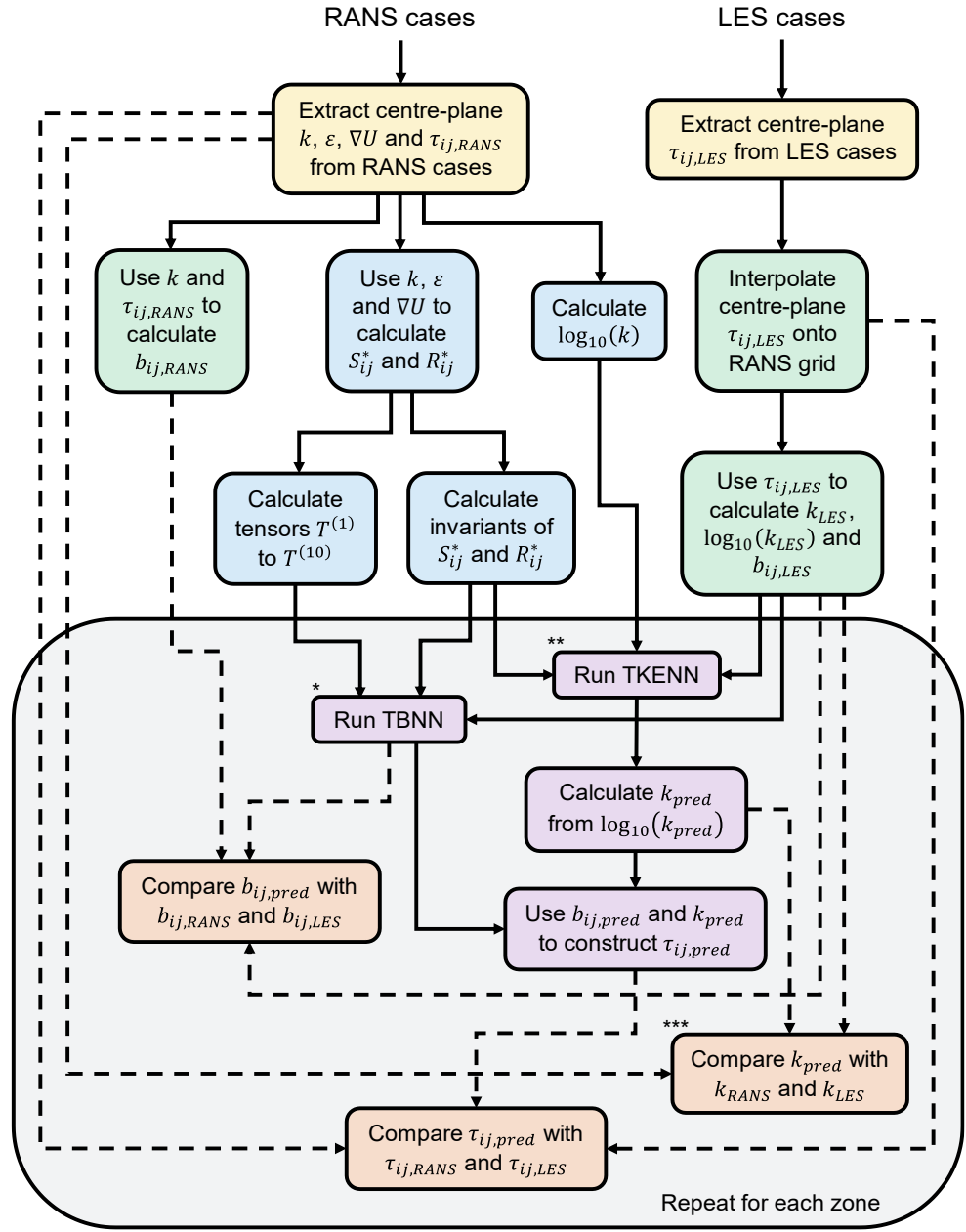

\* Run TBNN with invariants of $S_{ij}^*$ and $R_{ij}^*$ for the invariant input layer, tensors $T^{(1)}$ to $T^{(10)}$ for the tensor input layer and $b_{ij,LES}$ as the truth data to give prediction $b_{ij,pred}$
\*\* Run TKENN with $\log_{10}(k)$ and invariants of $S_{ij}^*$ and $R_{ij}^*$ for the input layer, and $\log_{10}(k_{LES})$ as the truth data to give prediction $\log_{10}(k_{pred})$
\*\*\* $k$ is equivalent to $k_{RANS}$ but $k_{RANS}$ has been written here for clarity

**Figure 6** The machine learning workflow deployed in the present modelling.

## 2D Datasets and Machine Learning Procedures

After calculating the input and output features for all Reynolds numbers, three datasets were formed from them for the ML procedures of training, validation and testing, and the training dataset was divided into batches. During TBNN or TKENN training, a batch of input features are inserted into the input layer, and the NN output is compared with the respective truth value provided by LES, thereby allowing a training error to be calculated. This error is fed back to each node in the hidden layers via backpropagation. By performing gradient descent on the errors, the weights and biases of the nodes are updated under a learning rate (LR), which allows the NN to become a more accurate mapping between the input and output features of the training dataset. This process is repeated for every training batch and can be performed repeatedly over the training dataset, where one full pass is known as an epoch. After an initial number of epochs, a ML model can be tasked with predicting the validation dataset. This enables early stopping of the training to prevent overfitting if a criterion is met. In this work, the NNs were queried on the validation dataset after every 4 epochs to obtain a validation error, and early stopping was triggered if the average of the latest three validation errors was greater than the three before. For TBNN, the validation error took the following form of mean squared error (MSE) [12]:

$$MSE_{TBNN} = \frac{1}{9N_s} \sum_{s=1}^{N_s} \sum_{i=1}^{3} \sum_{j=1}^{3} (b_{ij,pred} - b_{ij,LES})^2 \qquad (4)$$

where, $N_s$ is the number of data samples in the validation dataset, $b_{ij,pred}$ is $b_{ij}$ predicted by the TBNN (on the validation dataset when performing validation) and $b_{ij,LES}$ is ground-truth $b_{ij}$ from LES. The TKENN validation error took the following MSE adapted from Yin et al. [25] instead:

$$MSE_{TKENN} = \frac{1}{N_s} \sum_{s=1}^{N_s} (Y_{pred} - Y_{LES})^2 \qquad (5)$$

where, $Y_{pred}$ is the predicted TKE in logarithm form $\left(= \log_{10}(k_{pred})\right)$ and $Y_{LES} = \log_{10}(k_{LES})$.

Data from Re = [1800, 4500 and 5400] cases were allocated for training and validation. The selection of data from one of these cases as the validation dataset was rotated during hyperparameter tuning to facilitate three-fold cross-validation. Hyperparameters are parameters of a ML model whose values can influence the predictive capability [45]. The most significant ones in NNs are the number of hidden layers and nodes, LR, activation functions and batch size. Optimal hyperparameter values of the non-zonal TBNN and TKENN found from two grid searches are shown in **Table 1**. More information regarding the hyperparameter tuning process and other sets of hyperparameters that were highly ranked are detailed in Appendix B. The zonal TBNNs and TKENNs were run afterwards with the optimal hyperparameters of the non-zonal TBNN and TKENN respectively to demonstrate the merits of the zonal approach in comparison to the best performing non-zonal models. This involved all NNs being trained on the Re = 1800 and 5400 cases and validated with Re = 4500.

**Table 1** Optimal hyperparameters

| Hyperparameter | Non-zonal TBNN | Non-zonal TKENN |
| --- | --- | --- |
| Number of hidden layers | 2 | 5 |
| Number of hidden nodes per hidden layer | 50 | 10 |
| Initial learning rate | 0.01[*] | 0.01[*] |
| Activation functions | Swish | ELU |
| Batch size | 256 | 256 |

[*] with exponential LR scheduler factor of $\gamma = 0.98$

All ML procedures were implemented in the ML framework PyTorch, version 1.13. For all NNs, node weights were initialized with He initialization as the activation functions used were variants of ReLU [46-49]. During training, Eq. (4) and (5) with $N_s$ = number of data samples per training batch were used as the loss functions. The Adam optimizer [50] was chosen and the exponential LR scheduler was included with an initial LR of 0.01 to progressively reduce LR based on the number of elapsed epochs.

**2E   Zonal Approach based on Divide and Conquer (DaC) Technique**

In DaC approaches, a problem is collectively solved by multiple sub-problems, with each responsible for solving or predicting a certain part of the original problem [32]. A clear example of DaC in traditional CFD is in parallel computation of CFD cases. DaC is also used in traditional ML – for instance in (i) mixture of experts (MoE), where the problem space is divided into multiple regions and each region is predicted by one of several models that is most optimized for it and in (ii) decision trees, where each branch split can be considered a sub-model for solving a sub-problem [51, 52]. In the ML examples, DaC is used because when deployed for the intended purpose, the predictions that the sub-models collectively give are usually more accurate than when one model is used across the entire problem space. This stems from each sub-model being locally optimized for their respective sub-region, and each sub-region having a distinct behaviour or relationship between inputs and outputs [52]. Locally optimized models not based on DaC are also used in traditional CFD, such as empirical wall functions, the k-ω SST model and the use of porous zone models.

We drew inspiration from the DaC technique and locally optimized models to develop a zonal ML approach for improving RANS predictions. Having identified that many ML models in the literature cannot improve RANS-predicted Reynolds stress (or its projections) well in more than one region of flow physics, we propose training one ML model per region, and all the ML models will also be validated and tested on data from their respective regions. This leaves an important consideration: how should the domain be partitioned? It is clear that in the CFD examples of locally optimized models, sub-models are used according to regions of distinct behaviour in the output quantity, such as velocity. In the ML examples, the chosen number of expert models in MoE and tree splits are governed by the output prediction classes or values. Therefore, we suggest dividing the flow domain into regions that have similar output values or input-output relationship in the to-be-deployed ML models. Given that TBNN and TKENN predict $b_{ij}$ and $k$ respectively, we chose non-dimensional TKE, $\alpha = k/U_b^2$ as a zonal scalar indicator to partition the flow domain. Whilst $U_b$ is predefined, $k$ should be from the RANS results as $\alpha$ must be calculated in the absence of LES data.

Prior to running the zonal NNs, values of $\alpha$ and the number of zones must be chosen to partition the pre-processed RANS and LES data. In the present work, the data was divided into two zones to reduce the zonal partitioning complexity in this introduction of the zonal approach and to demonstrate its merits even with the minimum number of zones. We aimed to represent laminar and low turbulent near-wall regions with zone 1, and the highly turbulent region caused by flow separations with zone 2. It was recognized that the highly turbulent region in flows over a solid block could be easily identified as the region enclosed by high TKE gradient, as shown in **Figure 4(a)**. Therefore, cell-centre locations where $\alpha < 0.1$ were binned in zone 1 and the remaining locations were binned in zone 2, as this value of $\alpha$ enabled the shape of zone 2 across different Reynolds numbers to be very similar to their highly turbulent region. The zones are shown in **Figure 7** for all Reynolds numbers.

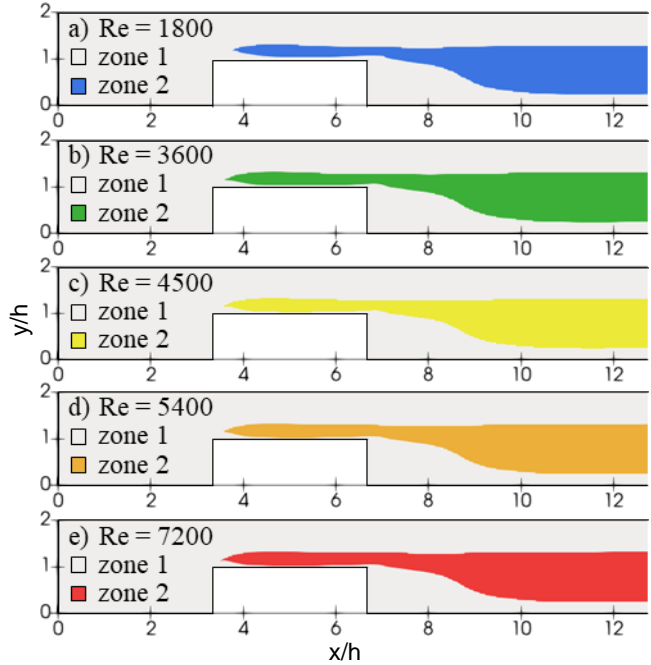

**Figure 7** Zone 1 (grey) and zone 2 (colored) for different Reynolds number cases.

**Figure 7** shows consistency in the shape of the zones across Reynolds numbers. This illustrates the Reynolds number independence of $\alpha$ arising from the normalization on $k$, which ensures regions with similar levels of turbulence relative to their bulk velocity in different Reynolds number cases can be identified as the same type of zone. Approximately 55,000 and 25,000 cell-centre locations were binned in zone 1 and 2 respectively for each Reynolds number. After partitioning the data, the ML workflow stages in the shaded part of **Figure 6** were repeated for each zone. For example with two zones demonstrated in this work, one TBNN and TKENN was trained, validated, and tested on data belonging to zone 1, and one TBNN and TKENN was trained, validated, and tested on data belonging to zone 2.

## 3 Results

After training and validating all NNs, their predictive capability was assessed by deploying them on the testing dataset, which consisted of data from Reynolds numbers inside and outside the training case range, *i.e.*, Re = 3600 and 7200. Approximately 20 CPU hours was required to train and validate the TBNNs and 2 CPU hours for the TKENNs, then only 10 minutes was required for testing the NNs. The test predictions given by the zonal TBNNs and TKENNs were combined into the full domain format. A spatial Gaussian filter was applied on these recombined zonal NN predictions to reduce discontinuity in the results at the border between the two zones, and to reduce noise in the NN predictions. Only plots of the normal streamwise (denoted subscript 11) and streamwise-perpendicular shear (denoted subscript 12) components are shown for $b_{ij}$ and constructed $\tau_{ij}$ predictions. This is because RANS prediction of $\tau_{11}$ gives the greatest discrepancy out of all normal components compared to LES, and thus contributes most to TKE inaccuracy, while $\tau_{12}$ is known to contribute significantly to separation bubble size [53].

### 3A  TBNN Predictions

Contours of $b_{11}$ above and downstream of the block for the Re = 3600 test case predicted by the non-zonal and zonal TBNNs are shown in **Figures 8(a)-(d)**. It is clear that both the non-zonal and zonal predictions are more accurate compared to RANS in all regions shown, including the near-wall, free-stream, and recirculation regions, which demonstrates the merits of TBNN in general. These improvements are quantitatively represented in **Table 2**, where RANS, non-zonal and zonal have been compared with LES as ground-truth to give MSEs. Additionally, the accuracy improvement given by these predictions over RANS (= $(MSE_{RANS} - MSE_{pred})/MSE_{RANS} \times 100\%$) denoted as RAI are shown. A perfectly predicted field equivalent to the LES result would give 100% RAI, whereas a predicted field with the same MSE as the RANS result would give 0% RAI. If the predicted field gives a higher MSE than the RANS result, then RAI becomes negative. Significant $b_{11}$ improvements over RANS prediction are recorded for the non-zonal and zonal approaches, with both leading to approximately 70% increase in $b_{11}$ accuracy for the whole flow domain. These results also show that zonal performed slightly better than non-zonal by 2% and 14% in zone 1 and 2 respectively, giving rise to an accuracy improvement of 3% over non-zonal in the whole domain. Both the non-zonal and zonal approaches produced a value of 0.13 for the mean $b_{11}$ across the entire domain, which was significantly more accurate than RANS which gave -0.01, given that the mean value from LES was 0.15.

To better illustrate the improvements by the zonal approach across both zone 1 and 2, line plots of $b_{11}$ at streamwise positions x/h = 4, 5, 8 and 10 are shown in **Figure 9**. These positions were chosen as lines perpendicular to them pass through the highly turbulent regions caused by the two separated flows. At all streamwise positions in **Figure 9**, the profile predicted by zonal follows the trend of LES closer than RANS and non-zonal, especially in the highly turbulent regions. However, there is still room for improvement in the upper-wall near-wall flow and free stream. These regions are part of zone 1 along with the laminar region upstream of the block. It is believed that further partitioning zone 1 into these regions of flow physics would lead to more accurate locally optimized models. Regardless, the current zonal prediction is still more accurate compared to non-zonal there, especially at x/h = 8 and 10 where LES gives a maximum value of 0.46, while zonal and non-zonal predicts 0.23 and 0.19 respectively.

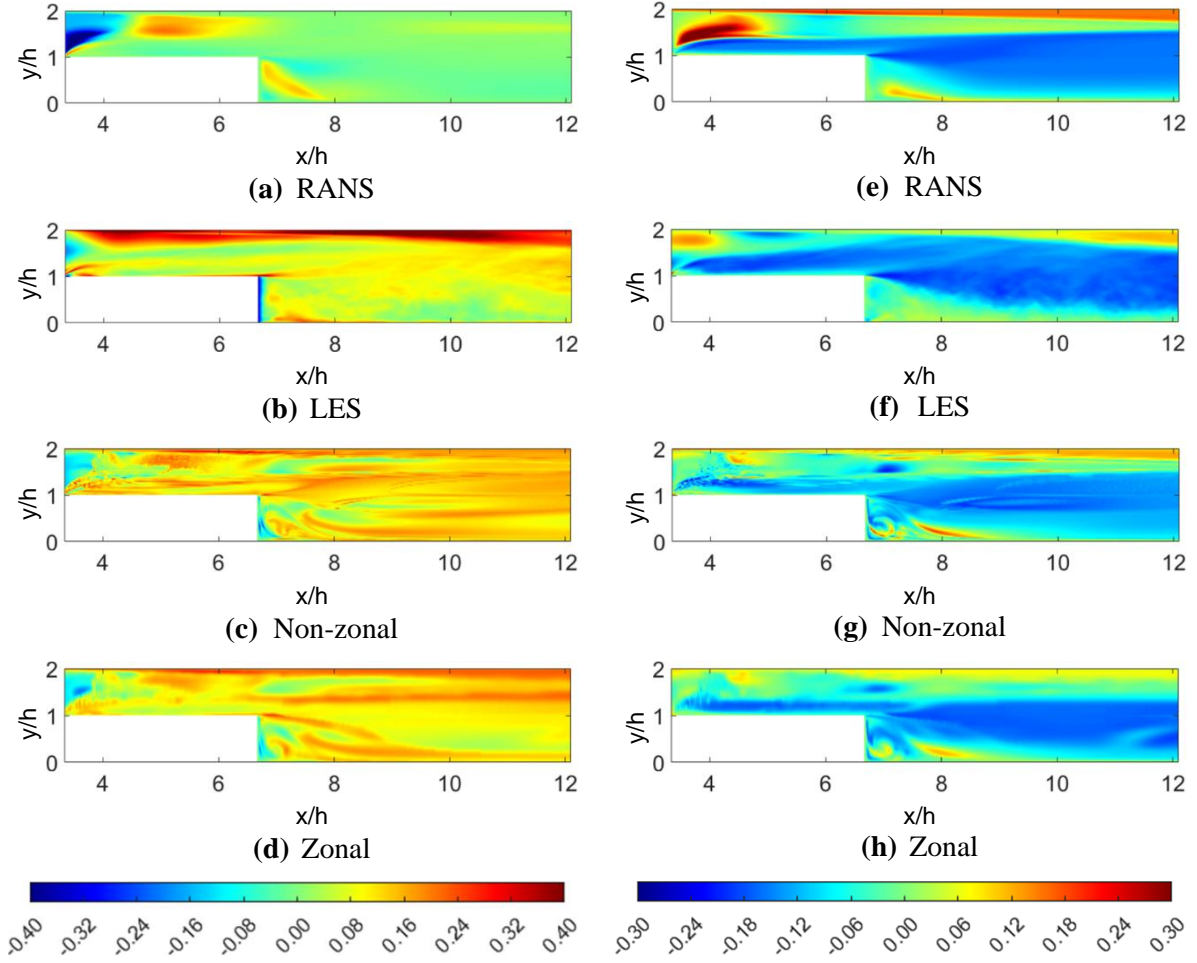

**Figure 8** Contours of **left:** $b_{11}$ and **right:** $b_{12}$ prediction results for Re = 3600 test case.

Contours of $b_{12}$ predictions for the Re = 3600 test case are shown in **Figures 8(e)-(h)**, and their MSE statistics are given in **Table 2**. Although the non-zonal and zonal approaches improved RANS $b_{12}$ accuracy by 42% and 62% respectively in the whole domain, the non-zonal approach reduced the accuracy of RANS by 65% in zone 2. This can be observed in **Figure 8(g)**, where the non-zonal result clearly underpredicts the magnitude of $b_{12}$ in the highly turbulent regions. In contrast, the zonal approach improved RANS prediction by 70% in zone 2, while also improving zone 1 accuracy by 62%. The improvement given by the zonal approach is also evident in the mean $b_{12}$ results across the whole domain, in which non-zonal and zonal produced a mean of -0.043 and -0.054 respectively – with the latter much closer to -0.061 given by LES. As the shape of zone 2 closely matches the region occupied by high magnitude negative $b_{12}$ in the LES result of **Figure 8(f)**, it suggests that the zonal TBNN for zone 2 was better optimized in training to model these high magnitude negative values compared to the non-zonal TBNN. Data from zone 1 must have influenced training of the non-zonal TBNN significantly, which indicates that the physics of $b_{12}$ in zone 2 are very different compared to the rest of the domain.

This result strongly supports our claim that the zonal approach should be applied in cases containing multiple flow physics and the justification for using non-dimensional TKE, $\alpha$ to partition the domain.

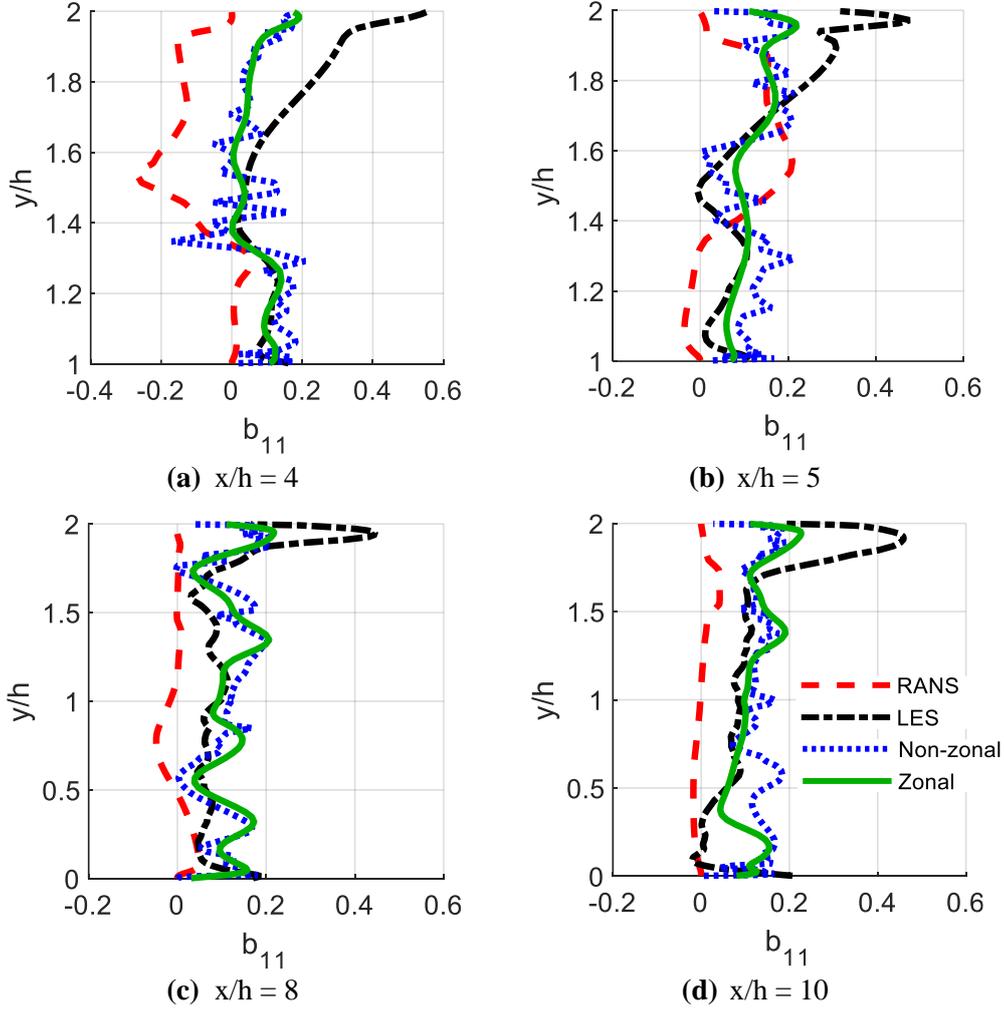

**Figure 9** Line plots of $b_{11}$ at different streamwise positions for Re = 3600 test case.

Line plots of $b_{12}$ predictions for the Re = 3600 case at the same streamwise stations as **Figure 9** are shown in **Figure 10**. Once again, the zonal profile follows the trend of LES more closely than RANS and non-zonal, especially in the wake downstream of the block at x/h = 8 and 10 where the zonal profile consistently matches the minimum given by LES. Although zonal gives a closer prediction to LES compared to non-zonal in the free stream and near the upper wall, there is still room for improvement here. The discussion on the $b_{12}$ contour plots of zone 2 reinforces the suggestions made in the analysis of the $b_{11}$ line plots – that partitioning zone 1 according to its regions of flow physics may give rise to more accurate zonal predictions in the free stream and flow near the upper wall as a result of the zonal TBNNs being more optimized for their individual region.

**Table 2** quantitatively shows the RANS accuracy improvement given by the non-zonal and zonal approaches for components $b_{22}$ and $b_{33}$. The prediction by zonal was more accurate than non-zonal for both components. While non-zonal gave 66% and 42% accuracy improvement over RANS for whole domain $b_{22}$ and $b_{33}$ respectively, the zonal prediction gave 74% and 46% improvement. $b_{ij}$ predictions for the Re = 3600 case given by the zonal TBNNs have therefore been shown to be more accurate on average compared to the non-zonal TBNN for all significant components of $b_{ij}$.

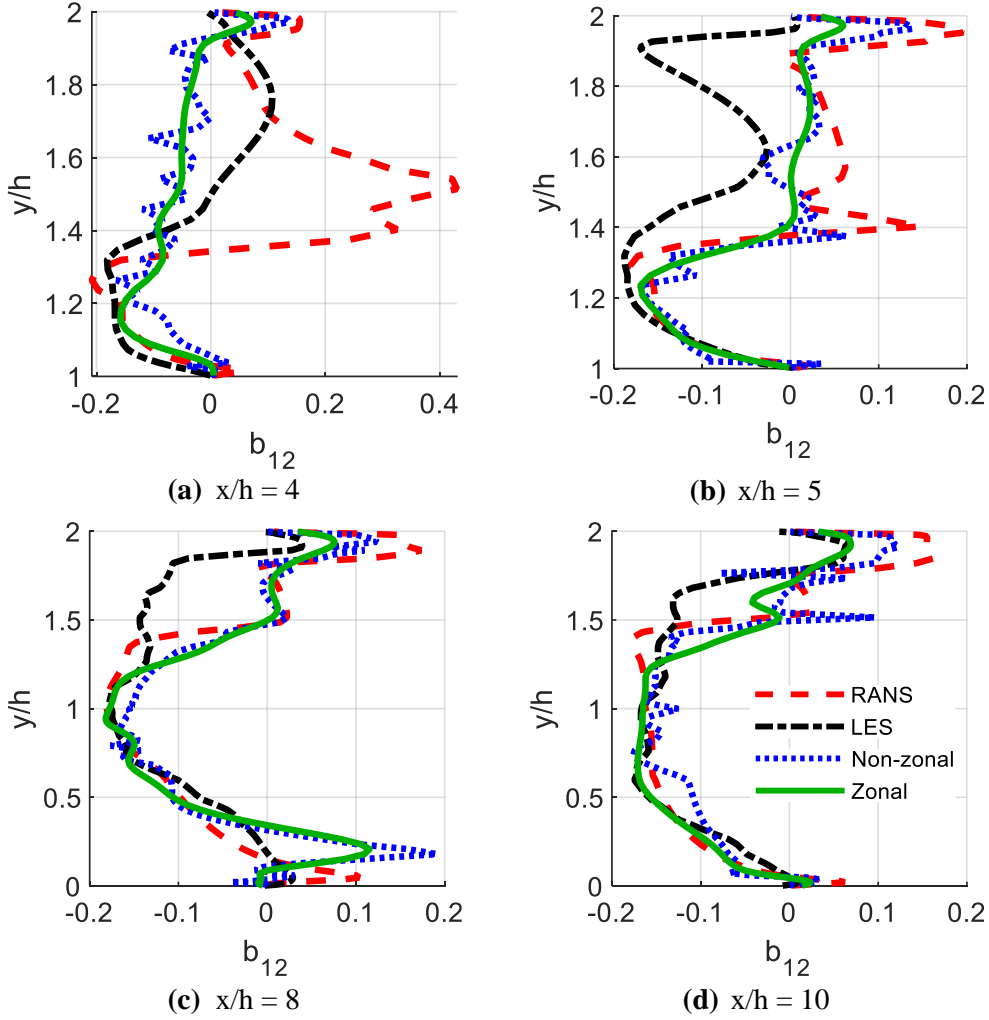

**Figure 10** Line plots of $b_{12}$ at different streamwise positions for Re = 3600 test case.

**Table 2** Anisotropy $b_{ij}$ Mean Squared Error with LES Data as Ground Truth for Re = 3600

| $b_{ij}$ component | Method | MSE compared to LES | | |
| --- | --- | --- | --- | --- |
| | | Whole domain | Zone 1 | Zone 2 |
| $b_{11}$ | RANS | $5.0\times10^{-2}$ | $6.7\times10^{-2}$ | $1.2\times10^{-2}$ |
| | Non-zonal (RAI %)[*] | $1.5\times10^{-2}$ (69%) | $2.1\times10^{-2}$ (68%) | $2.1\times10^{-3}$ (83%) |
| | Zonal (RAI %) | $1.4\times10^{-2}$ (72%) | $2.0\times10^{-2}$ (70%) | $3.7\times10^{-4}$ (97%) |
| $b_{12}$ | RANS | $7.1\times10^{-3}$ | $9.9\times10^{-3}$ | $9.2\times10^{-4}$ |
| | Non-zonal (RAI %) | $4.1\times10^{-3}$ (42%) | $5.3\times10^{-3}$ (47%) | $1.5\times10^{-3}$ (-65%) |
| | Zonal (RAI %) | $2.7\times10^{-3}$ (62%) | $3.8\times10^{-3}$ (62%) | $2.8\times10^{-4}$ (70%) |
| $b_{22}$ | RANS | $4.0\times10^{-2}$ | $5.5\times10^{-2}$ | $5.7\times10^{-3}$ |
| | Non-zonal (RAI %) | $1.3\times10^{-2}$ (66%) | $1.8\times10^{-2}$ (67%) | $2.2\times10^{-3}$ (61%) |
| | Zonal (RAI %) | $1.0\times10^{-2}$ (74%) | $1.5\times10^{-2}$ (73%) | $3.5\times10^{-4}$ (94%) |
| $b_{33}$ | RANS | $2.0\times10^{-2}$ | $2.9\times10^{-2}$ | $2.5\times10^{-3}$ |
| | Non-zonal (RAI %) | $1.2\times10^{-2}$ (42%) | $1.7\times10^{-2}$ (41%) | $7.2\times10^{-4}$ (71%) |
| | Zonal (RAI %) | $1.1\times10^{-2}$ (46%) | $1.6\times10^{-2}$ (45%) | $2.6\times10^{-4}$ (89%) |

[*] Reynolds Accuracy Improvement (RAI) % = $(MSE_{RANS} - MSE_{pred})/MSE_{RANS} \times 100\%$.

Predictions of the entire $b_{ij}$ tensor can be visualized on a barycentric map (BM), as shown in **Figure 11(a)**. Introduced by Banerjee et al. [54], the BM is a triangular Euclidean domain that contains all realizable states of turbulence. Each corner represents a limiting state, namely (i) one-component, where fluctuations only exist in one direction, (ii) axisymmetric two-component, where fluctuations exist in two directions with equivalent magnitude and (iii) isotropic three-component [55]. In the BM, any realizable state of turbulence can be represented by a convex combination of these three limiting states. To plot $b_{ij}$ on a BM, the eigenvalues of $b_{ij}$ must be sorted in descending order, and functions of them are used to calculate coefficients, which are combined to give the (x, y) barycentric coordinates. Readers are referred to Banerjee et al. [54] for more information on the BM plotting process.

Given that the highly turbulent regions caused by the two flow separations dominate the fluid dynamics in flows over a solid block, the BM was plotted across them in **Figures 11** and **12**. This entailed $b_{ij}$ results from RANS, LES, non-zonal and zonal approaches being extracted along the red dashed lines in **Figures 11(b)** and **12(b)**, and plotted on the BMs in **Figures 11(a)** and **12(a)**, with zoomed-in views provided in **Figures 11(c)** and **12(c)**. The marker shapes increase in size with distance along the red dashed lines from left to right. **Figures 11(c)** and **12(c)** show that the anisotropy given by the LES result trends from one-component towards the two- and three-component states with increasing distance along the sampling lines. This is because the streamwise component of Reynolds stress $\overline{u'^2}$ dominates the other two normal components just after the separation point. The other two components increase gradually with distance downstream afterwards, leading towards more isotropic turbulence. These results agree with the study by Jadidi et al. [56], where a porous block was used instead of a solid one in their LES computations. The average direction given by the mean vector between sampling points in **Figures 11**(c) and **12**(c) shows that this trend is predicted by the zonal approach, while non-zonal fails to predict this above the block, and RANS in both BMs.

Average BM coordinates are also plotted in **Figures 11**(c) and **12**(c) for the four methods. **Figure 11**(c) shows that the average zonal coordinate is approximately 2.5 times further away from the LES coordinate than non-zonal. However, it is believed that the points on the right side of **Figure 11**(c) representing the locations closest to the separation point have skewed the averages. Regardless, it is clear that the zonal approach predictions can be improved here where zone 1 and 2 meet, whilst the results given by the zonal and non-zonal approaches match well with LES further downstream. **Figure 12**(c) shows that the distance between the average zonal and LES coordinate is 45% of the distance between non-zonal and LES. Given that the entire sampling line lies within zone 2, this result shows that the zonal approach is able to predict the anisotropic state more accurately than RANS and non-zonal in locations firmly in the zones. This reinforces that the zonal TBNN for zone 2 was more optimized at predicting $b_{ij}$ in the highly turbulent regions compared to the non-zonal TBNN.

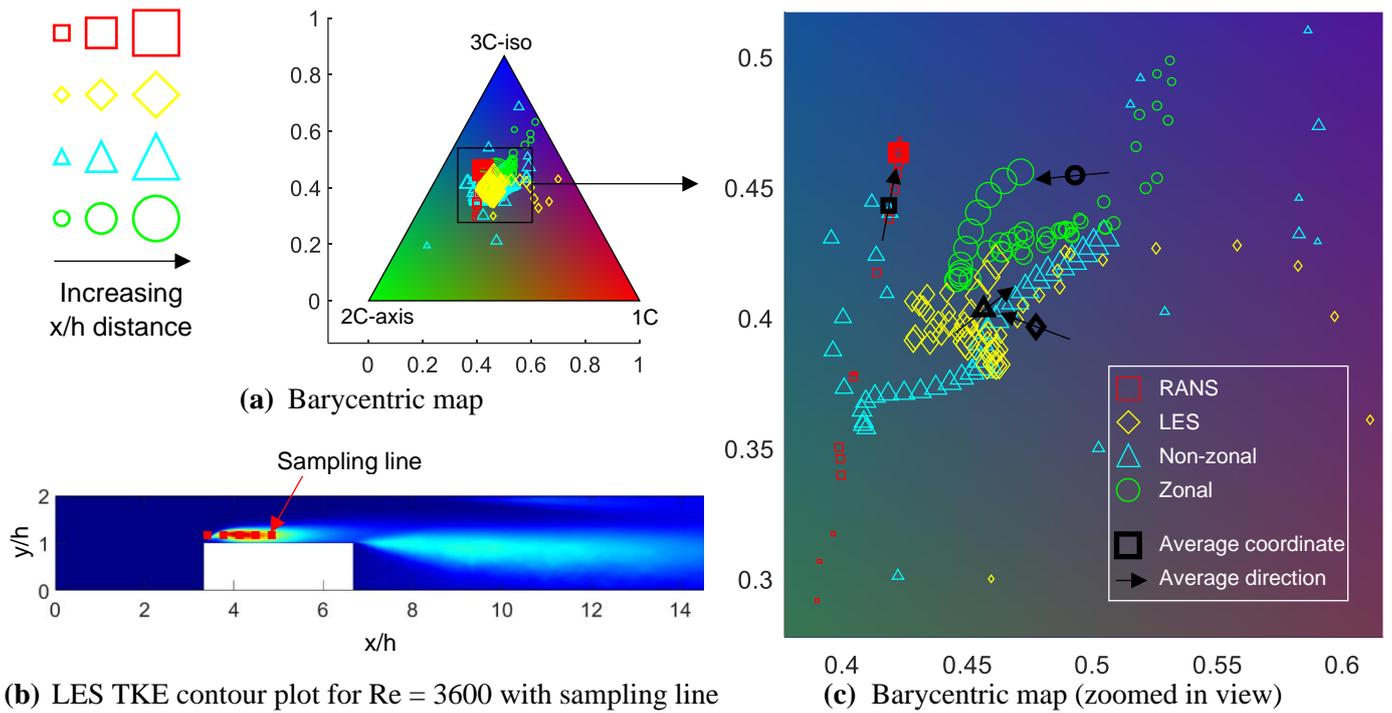

**Figure 11** Barycentric map showing anisotropy predictions in the highly turbulent region above the block for Re = 3600 test case ($U_b$ = 2 m/s). The anisotropy was sampled along the red dashed line in **(b)** where there is high TKE. Symbol size increases with distance from left to right along the red dashed line.

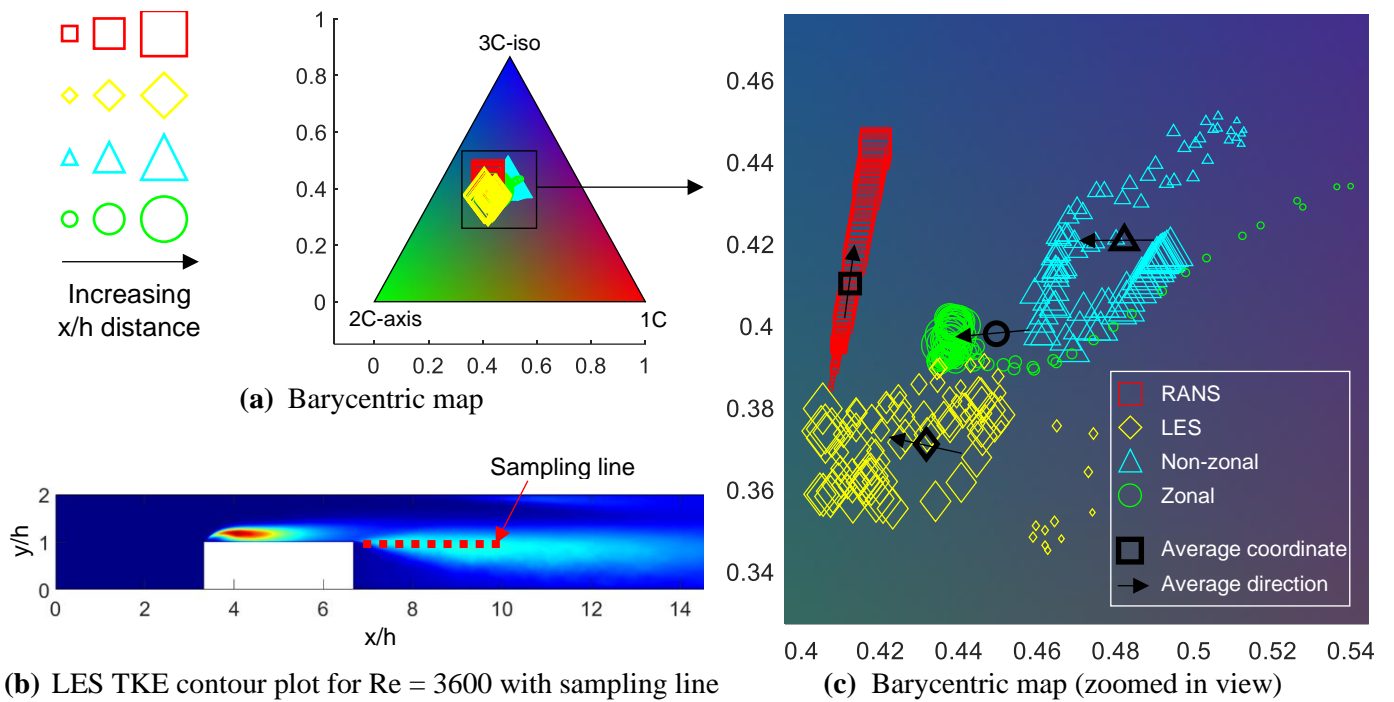

**Figure 12** Barycentric map showing anisotropy predictions in the highly turbulent region downstream of the block for Re = 3600 test case ($U_b$ = 2 m/s). The anisotropy was sampled along the red dashed line in **(b)** where there is high TKE. Symbol size increases with distance from left to right along the red dashed line.

Due to the Reynolds number invariance of $b_{ij}$ [17], the predictions given by the non-zonal and zonal TBNNs for the Re = 7200 test case were very similar to those shown in **Figure 8**. Both non-zonal and zonal predicted $b_{11}$ more accurately compared to RANS across all regions, as recorded in **Table 3**. Although non-zonal improved RANS $b_{11}$ MSE by 57%, 86% and 60% in zone 1, zone 2 and the whole domain respectively, the zonal approach gave greater improvements of 61% in zone 1 and 96% in zone 2, leading to 64% in the whole domain overall. Similar to the previous test case, $b_{12}$ was underpredicted again in zone 2 by the non-zonal TBNN for Re = 7200. This resulted RANS $b_{12}$ MSE increasing by 127% for zone 2 as recorded in **Table 3**. In contrast, the zonal approach prediction improved RANS $b_{12}$ accuracy by 43% in zone 2 and 55% in zone 1, leading to 55% improvement across the whole domain, which amounted to 24% higher than the non-zonal TBNN. The biggest discrepancies between the zonal and LES results for this test case remained in the free stream and near-wall region by the upper wall, which may be remedied by further partitioning zone 1 into its regions of flow physics, as given in the discussion of the previous test case.

**Table 3** Anisotropy $b_{ij}$ Mean Squared Error with LES Data as Ground Truth for Re = 7200

| $b_{ij}$ component | Method | MSE compared to LES | | |
| --- | --- | --- | --- | --- |
| | | Whole domain | Zone 1 | Zone 2 |
| $b_{11}$ | RANS | $4.8 \times 10^{-2}$ | $6.4 \times 10^{-2}$ | $1.4 \times 10^{-2}$ |
| | Non-zonal (RAI %)[*] | $1.9 \times 10^{-2}$ (60%) | $2.7 \times 10^{-2}$ (57%) | $2.0 \times 10^{-3}$ (86%) |
| | Zonal (RAI %) | $1.7 \times 10^{-2}$ (64%) | $2.5 \times 10^{-2}$ (61%) | $6.1 \times 10^{-4}$ (96%) |
| $b_{12}$ | RANS | $6.5 \times 10^{-3}$ | $9.3 \times 10^{-3}$ | $8.4 \times 10^{-4}$ |
| | Non-zonal (RAI %) | $4.5 \times 10^{-3}$ (31%) | $5.8 \times 10^{-3}$ (38%) | $1.9 \times 10^{-3}$ (-127%) |
| | Zonal (RAI %) | $3.0 \times 10^{-3}$ (55%) | $4.2 \times 10^{-3}$ (55%) | $4.8 \times 10^{-4}$ (43%) |
| $b_{22}$ | RANS | $3.6 \times 10^{-2}$ | $5.1 \times 10^{-2}$ | $5.7 \times 10^{-3}$ |
| | Non-zonal (RAI %) | $1.2 \times 10^{-2}$ (67%) | $1.6 \times 10^{-2}$ (68%) | $2.3 \times 10^{-3}$ (59%) |
| | Zonal (RAI %) | $8.6 \times 10^{-3}$ (76%) | $1.3 \times 10^{-2}$ (75%) | $4.3 \times 10^{-4}$ (92%) |
| $b_{33}$ | RANS | $1.7 \times 10^{-2}$ | $2.3 \times 10^{-2}$ | $3.2 \times 10^{-3}$ |
| | Non-zonal (RAI %) | $1.8 \times 10^{-2}$ (-9%) | $2.7 \times 10^{-2}$ (-15%) | $9.2 \times 10^{-4}$ (71%) |
| | Zonal (RAI %) | $1.7 \times 10^{-2}$ (0%) | $2.5 \times 10^{-2}$ (-5%) | $4.2 \times 10^{-4}$ (87%) |

[*] Reynolds Accuracy Improvement (RAI) % = $(MSE_{RANS} - MSE_{pred})/MSE_{RANS} \times 100\%$

## 3B  TKENN Predictions

Contours of TKE for the Re = 3600 test case predicted by non-zonal and zonal TKENNs are shown in **Figures 13(c)** and **13(d)** respectively. Curves of uncharacteristically low TKE downstream of the block can be observed in **Figure 13(c)**, which present a source of inaccuracy in the non-zonal prediction. These were found to be caused by a non-unique mapping between the input features and the output $\log_{10}(k)$ [17]. Non-unique mapping occurs when a ML model is trained to map the same input values to different output values. In this case, points along these curves share the same input values as other points in the domain but have different $\log_{10}(k_{LES})$ true output values. This led to the non-zonal approach worsening the RANS $k$ MSE in both zones as shown in **Table 4**, resulting in a 12% reduction in RANS accuracy across the whole domain.

The zonal approach can overcome non-unique mappings by allocating the conflicting points to different zones, so that their data does not appear in the same training dataset. This can be easily achieved, as these points usually occur in different regions of flow physics [17]. As a result, the curves of low TKE are absent in zone 2 of the zonal approach prediction given in **Figure 13(d)**. Some remain downstream of the block in the wake region of zone 1, due to non-unique mapping between these locations and the near-wall flow by the upper wall, which presents another motivation to divide zone 1 further into its regions of flow physics. Regardless, the simple zonal partitioning employed in this work enabled the shape of zone 2 above the block to be similar to the high TKE contour shape, allowing the zone 2 TKENN to be better optimized at predicting TKE there. Due to this and overcoming the non-unique mapping in some locations, the zonal approach improved RANS accuracy in both zone 1 and 2 by 5% and 24% respectively, resulting in 16% improvement across the entire domain as shown in **Table 4**.

Contours of TKE for the non-zonal and zonal predictions for the Re = 7200 case are shown in **Figures 13(g)** and **13(h),** respectively. There are some similarities with the Re = 3600 test results in **Figures 13(c)** and **13(d)**. For example, the high TKE region shape above the block was better predicted by the zonal approach due to the shape of zone 2. Secondly, two low TKE curves can still be observed in the non-zonal prediction downstream of the block as shown in **Figure 13(g)**. In contrast, these are not present in the zonal prediction given in **Figure 13(h)**, which may be due to the zonal partitioning of training data belonging to conflicting points as discussed in the TKENN results of the previous test case. **Figures 13(g)** and **13(h)** show that the non-zonal and zonal approaches underpredict TKE above the block, which did not occur in the Re = 3600 test case results. This shows that $\log_{10}(k)$ is not Reynolds number invariant and therefore should not be assigned as the output of TKENNs to guarantee good extrapolation performance. Setting $\log_{10}(k/U_b^2)$ as the output instead may result in Reynolds number independence in a trained TKENN. Despite this, the zonal approach was still able to improve RANS accuracy by 34% in zone 1 and 44% in zone 2, leading to 40% improvement over RANS across the whole domain as shown in **Table 4**. This corresponded to 40%, 27% and 32% more than the non-zonal approach in zone 1, zone 2 and the whole domain respectively.

Table 4 TKE Mean Squared Error with LES Data as Ground Truth

| Reynolds number | Method | MSE compared to LES | | |
| --- | --- | --- | --- | --- |
| | | Whole domain | Zone 1 | Zone 2 |
| 3600 | RANS | 0.036 | 0.022 | 0.067 |
| | Non-zonal (RAI %)* | 0.041 (-12%) | 0.023 (-3%) | 0.079 (-18%) |
| | Zonal (RAI %) | 0.030 (16%) | 0.021 (5%) | 0.051 (24%) |
| 7200 | RANS | 0.529 | 0.306 | 0.992 |
| | Non-zonal (RAI %) | 0.487 (8%) | 0.325 (-6%) | 0.823 (17%) |
| | Zonal (RAI %) | 0.316 (40%) | 0.201 (34%) | 0.556 (44%) |

* Reynolds Accuracy Improvement (RAI) % = $(MSE_{RANS} - MSE_{pred})/MSE_{RANS} \times 100\%$

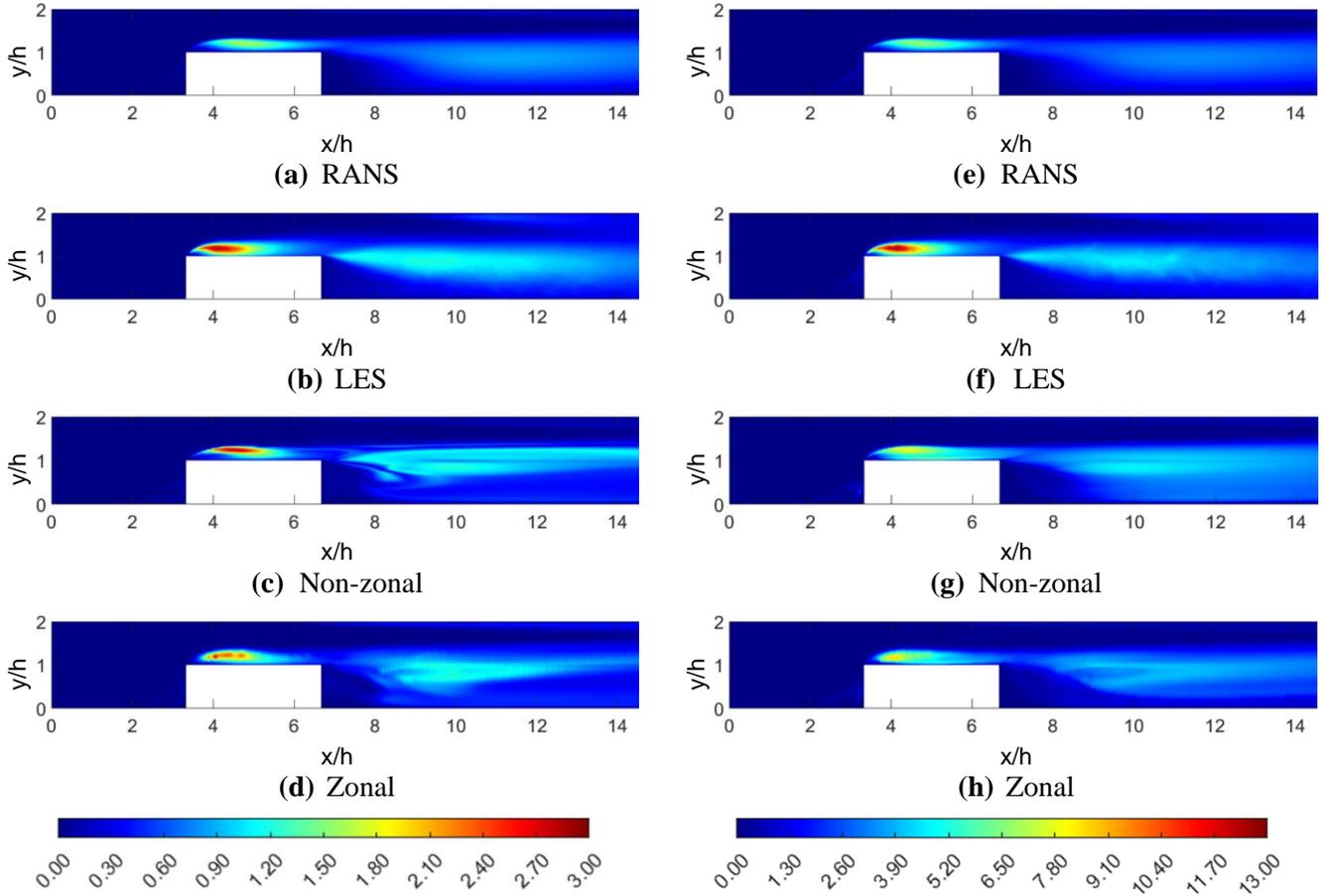

**Figure 13** Contours of TKE (m$^2$/s$^2$) prediction results for **left:** Re = 3600 and **right:** Re = 7200 test cases.

### 3C  Constructed Reynolds Stress for Re = 3600 Test Case

The $b_{ij}$ predictions from TBNN and $k$ predictions from TKENN were combined to give constructed Reynolds stress $\tau_{ij}$ using Eq. (2). The non-zonal $\tau_{ij}$ inference was constructed from the non-zonal TBNN and TKENN predictions, and the zonal $\tau_{ij}$ inference was constructed from the zonal TBNN and TKENN predictions. Contour plots of $\tau_{11}$ for the Re = 3600 test case given by the two approaches are shown in **Figures 14(c)** and **14(d)**. A comparison between **Figures 8** and **13** shows that the magnitude of predicted TKE is much greater than predicted $b_{ij}$ across much of the domain. Given this magnitude difference and Eq. (2), it is not surprising that the TKE predictions have a strong influence on the distribution of $\tau_{11}$. For example, the distribution of TKE predicted by the non-zonal approach in **Figure 13(c)** is almost identical to that of $\tau_{11}$ in **Figure 14(c)**, while the same observation can be made for the zonal predictions in **Figures 13(d)** and **14(d)**. Therefore for a posteriori analysis, TKE should be accurately predicted when TBNN is used, in order for accurate predictions of Reynolds stress to be propagated back into the RANS solver. Due to improvements in RANS accuracy made by the zonal NNs, a 38% increase in RANS accuracy was achieved by the zonal approach for the $\tau_{11}$ prediction across the whole domain as shown in **Table 5**. This amounted to 23% more than the non-zonal approach. **Figures 14(g)** and **14(h)** show contour plots of constructed $\tau_{12}$ by the two approaches for the Re = 3600 test case. Similar to $\tau_{11}$, the TKE predictions seem to strongly influence the distribution of constructed $\tau_{12}$ as well. Therefore, $\tau_{12}$ was also more accurately predicted by the zonal approach than the non-zonal models. RANS $\tau_{12}$ MSE worsened by 4% and 15% respectively in zone 1 and 2 in the non-zonal result, leading to an accuracy reduction of 12% across the whole domain as shown in **Table 5**. In contrast, the zonal approach improved RANS accuracy by 22% and 33% respectively for zone 1 and 2, resulting in 30% improvement across the whole domain.

**Table 5** Reynolds Stress Mean Squared Error with LES Data as Ground Truth for Re = 3600

| $\tau_{ij}$ component | Method | MSE compared to LES | | |
| --- | --- | --- | --- | --- |
| | | Whole domain | Zone 1 | Zone 2 |
| $\tau_{11}$ | RANS | $4.4\times10^{-2}$ | $2.6\times10^{-2}$ | $8.4\times10^{-2}$ |
| | Non-zonal (RAI %)* | $3.7\times10^{-2}$ (15%) | $2.4\times10^{-2}$ (4%) | $6.6\times10^{-2}$ (21%) |
| | Zonal (RAI %) | $2.7\times10^{-2}$ (38%) | $2.3\times10^{-2}$ (11%) | $3.7\times10^{-2}$ (56%) |
| $\tau_{12}$ | RANS | $4.0\times10^{-3}$ | $1.4\times10^{-3}$ | $9.5\times10^{-3}$ |
| | Non-zonal (RAI %) | $4.5\times10^{-3}$ (-12%) | $1.5\times10^{-3}$ (-4%) | $1.1\times10^{-2}$ (-15%) |
| | Zonal (RAI %) | $2.8\times10^{-3}$ (30%) | $1.1\times10^{-3}$ (22%) | $6.4\times10^{-3}$ (33%) |

* Reynolds Accuracy Improvement (RAI) % = $(MSE_{RANS} - MSE_{pred})/MSE_{RANS} \times 100\%$

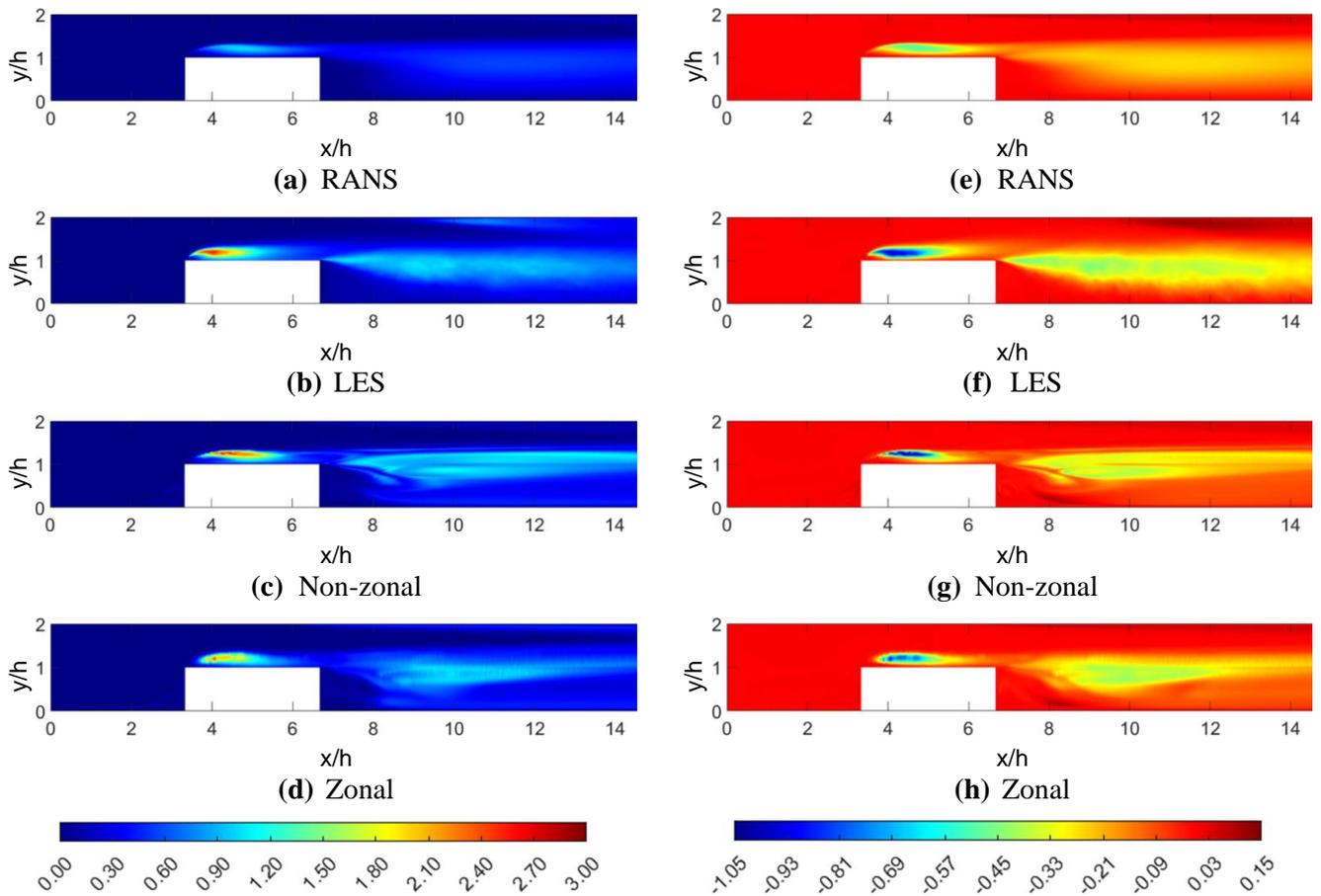

**Figure 14** Contours of constructed **left:** $\tau_{11}$ and **right:** $\tau_{12}$ results for Re = 3600 test case (m$^2$/s$^2$).

### 3D Constructed Reynolds Stress for Re = 7200 Test Case

**Figure 15** shows contours of constructed $\tau_{11}$ and $\tau_{12}$ given by the two approaches for the Re = 7200 test case. A comparison of these plots with **Figure 13** shows that similar to the constructed Reynolds stresses for the Re = 3600 test case, the predicted TKE strongly influences the distribution of $\tau_{11}$ and $\tau_{12}$ in this test case as well. The underprediction of TKE above the block in **Figure 13** has also led to $\tau_{11}$ and $\tau_{12}$ being underpredicted there. Regardless, both non-zonal and zonal approaches were able to improve RANS accuracy of Reynolds stress, as shown in **Table 6**. While the non-zonal approach improved RANS accuracy in the whole domain by 26% and 14% for $\tau_{11}$ and $\tau_{12}$ respectively, the zonal approach gave higher improvements of 61% for $\tau_{11}$ and 43% for $\tau_{12}$. These greater improvements were driven by the zonal NNs predicting more accurately than their non-zonal counterparts, as shown in **Tables 3** and **4**.

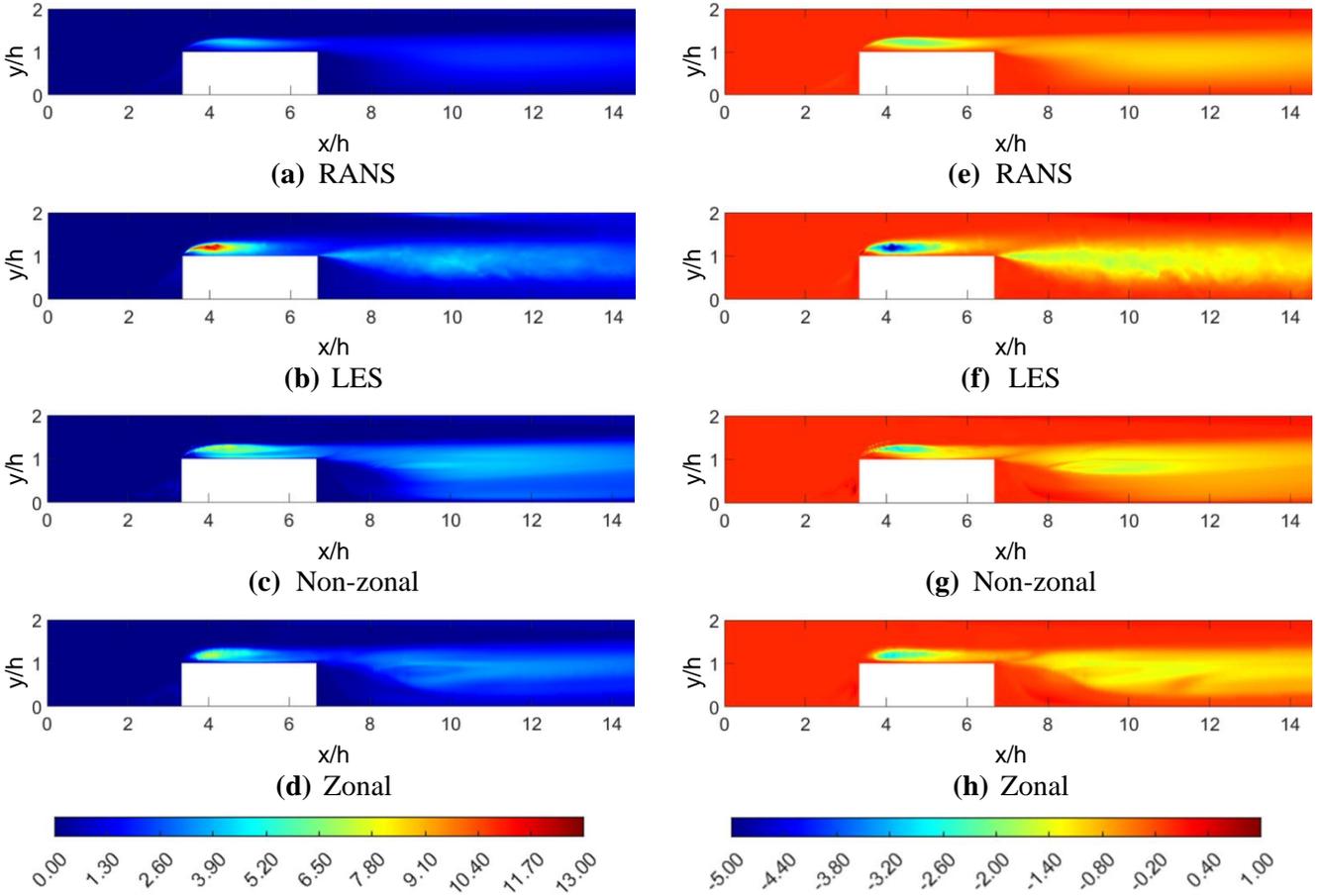

**Figure 15** Contours of constructed **left:** $\tau_{11}$ and **right:** $\tau_{12}$ results for Re = 7200 test case (m²/s²).

**Table 6** Reynolds Stress Mean Squared Error with LES Data as Ground Truth for Re = 7200

| $\tau_{ij}$ component | Method | MSE compared to LES | | |
|---|---|---|---|---|
| | | Whole domain | Zone 1 | Zone 2 |
| $\tau_{11}$ | RANS | 0.706 | 0.371 | 1.401 |
| | Non-zonal (RAI %)* | 0.520 (26%) | 0.366 (1%) | 0.838 (40%) |
| | Zonal (RAI %) | 0.273 (61%) | 0.203 (45%) | 0.419 (70%) |
| $\tau_{12}$ | RANS | 0.059 | 0.021 | 0.137 |
| | Non-zonal (RAI %) | 0.051 (14%) | 0.022 (-6%) | 0.109 (20%) |
| | Zonal (RAI %) | 0.034 (43%) | 0.017 (18%) | 0.068 (50%) |

* Reynolds Accuracy Improvement (RAI) % = $(MSE_{RANS} - MSE_{pred})/MSE_{RANS} \times 100\%$

# 4 Conclusion

A divide-and-conquer machine learning (ML) approach for RANS turbulence modelling was presented and applied to two ML models: the Tensor Basis Neural Network (TBNN) and a neural network (NN) for predicting turbulent kinetic energy (TKE) called TKENN. These models were deployed on cases of flow over a solid block, which contain different regions of flow physics including separated flows. It was shown that partitioning the flow domain into two zones according to its flow physics and training separate TBNNs and TKENNs on each one allowed the NNs to be highly optimized at predicting their respective zone. In every test case undertaken, the combined zonal prediction result was more accurate than RANS and the prediction given by the same NN trained on the entire domain.

The prediction accuracy of all Reynolds stress anisotropy $b_{ij}$ components given by the zonal TBNNs was higher in both zones than RANS and the non-zonal TBNN. Most notably, this led to the accuracy of $b_{12}$ being improved by at least 55% over RANS and 20% over the non-zonal TBNN prediction for the entire flow domain of the test cases. These improvements were attributed to one of the zones and the region of high magnitude negative $b_{12}$ having a similar shape. Therefore, the TBNN for this zone was more optimized at predicting the high magnitude negative $b_{12}$ region than the non-zonal TBNN. This finding shows that zonal ML models can perform well when the shape of the zones coincide with regions of similar output values or input-output relationship. Predictions of $b_{ij}$ plotted on barycentric maps also showed that the zonal approach was able to predict the states of anisotropy more accurately in the highly turbulent regions than RANS and the non-zonal TBNN.

The zonal TKENNs proposed in this work were shown to predict TKE more accurately in both zones than RANS and the non-zonal TKENN. This led to the zonal approach predicting TKE at least 28% more accurately than non-zonal, and 16% more than RANS on average for the entire flow domain of the test cases. The higher accuracy given by the zonal approach over non-zonal was attributed to the zone partitioning being able to eliminate non-unique mappings between TKENN inputs and output, and the shape of the high TKE region above the block being better predicted due to the zone responsible for predicting that region having a similar shape there. Lastly, combining the zonal TBNN and TKENN predictions to form constructed Reynolds stress showed that the zonal approach was able to predict this quantity more accurately than RANS and the non-zonal approach in both zones as well.

## Acknowledgements

This work was supported by the UK Engineering and Physical Sciences Research Council (EPSRC) [grant numbers EP/W033542/1 and EP/T012242/2]. The authors would like to acknowledge the assistance given by Research IT and the use of the Computational Shared Facility at The University of Manchester.

## Appendix A: Tensor Basis Neural Network

Pope [11] postulated that the Reynolds stresses $\tau_{ij}$ (and Reynolds stress anisotropy $b_{ij}$) are uniquely related to local rates of strain and local turbulent scalar quantities, provided that (i) the flow has a high enough Reynolds number such that all macroscales of turbulence are proportional, (ii) at locations far from a boundary, the boundary conditions affect the turbulent time and length scale only and (iii) there is negligible Reynolds stress transport and thus the rates of strain are nearly homogeneous. This unique relation is called the general effective-viscosity hypothesis (GEVH), and is a tensor polynomial expression that has a closed form consisting of a sum of 10 terms for three-dimensional flows:

$$\begin{aligned} b_{ij} = &\ g^{(1)} \underbrace{\boldsymbol{S}}_{\boldsymbol{T}^{(1)}} + g^{(2)} \underbrace{(\boldsymbol{SR} - \boldsymbol{RS})}_{\boldsymbol{T}^{(2)}} + g^{(3)} \underbrace{\left(\boldsymbol{S}^2 - \frac{1}{3}\boldsymbol{I}\{\boldsymbol{S}^2\}\right)}_{\boldsymbol{T}^{(3)}} + g^{(4)} \underbrace{\left(\boldsymbol{R}^2 - \frac{1}{3}\boldsymbol{I}\{\boldsymbol{R}^2\}\right)}_{\boldsymbol{T}^{(4)}} \\ &+ g^{(5)} \underbrace{(\boldsymbol{RS}^2 - \boldsymbol{S}^2\boldsymbol{R})}_{\boldsymbol{T}^{(5)}} + g^{(6)} \underbrace{\left(\boldsymbol{R}^2\boldsymbol{S} + \boldsymbol{SR}^2 - \frac{2}{3}\boldsymbol{I}\{\boldsymbol{SR}^2\}\right)}_{\boldsymbol{T}^{(6)}} \\ &+ g^{(7)} \underbrace{(\boldsymbol{RSR}^2 - \boldsymbol{R}^2\boldsymbol{SR})}_{\boldsymbol{T}^{(7)}} + g^{(8)} \underbrace{(\boldsymbol{SRS}^2 - \boldsymbol{S}^2\boldsymbol{RS})}_{\boldsymbol{T}^{(8)}} \\ &+ g^{(9)} \underbrace{\left(\boldsymbol{R}^2\boldsymbol{S}^2 + \boldsymbol{S}^2\boldsymbol{R}^2 - \frac{2}{3}\boldsymbol{I}\{\boldsymbol{S}^2\boldsymbol{R}^2\}\right)}_{\boldsymbol{T}^{(9)}} + g^{(10)} \underbrace{(\boldsymbol{RS}^2\boldsymbol{R}^2 - \boldsymbol{R}^2\boldsymbol{S}^2\boldsymbol{R})}_{\boldsymbol{T}^{(10)}} \end{aligned} \quad (A1)$$

$g^{(1)}$ to $g^{(10)}$ are scalar coefficients that are unknown functions of the following five non-dimensional mean strain rate $S_{ij}^*(=kS_{ij}/\varepsilon)$ and mean rotation rate $R_{ij}^*(=kR_{ij}/\varepsilon)$ invariants:

$$g^{(n)} = f(\{\boldsymbol{S}^2\}, \{\boldsymbol{R}^2\}, \{\boldsymbol{S}^3\}, \{\boldsymbol{R}^2\boldsymbol{S}\}, \{\boldsymbol{R}^2\boldsymbol{S}^2\}) \quad (A2)$$

where, $\{\boldsymbol{S}^2\} = S_{ik}^* S_{ki}^*$, $\{\boldsymbol{R}^2\boldsymbol{S}\} = R_{ik}^* R_{kl}^* S_{li}^*$, $\{\boldsymbol{R}^2\boldsymbol{S}^2\} = R_{ik}^* R_{kl}^* S_{lm}^* S_{mi}^*$ etc. in Einstein notation, $k$ is the turbulent kinetic energy (TKE), $\varepsilon$ is the TKE dissipation rate and $\boldsymbol{I}$ is Kronecker delta. The tensors in $\boldsymbol{T}^{(1)}$ to $\boldsymbol{T}^{(10)}$ are abbreviated as $\boldsymbol{S} = S_{ij}^*$, $\boldsymbol{SR} = S_{ik}^* R_{kj}^*$, $\boldsymbol{S}^2\boldsymbol{R} = S_{ik}^* S_{kl}^* R_{lj}^*$, $\boldsymbol{S}^2\boldsymbol{R}^2 = S_{ik}^* S_{kl}^* R_{lm}^* R_{mj}^*$ and $\boldsymbol{RS}^2\boldsymbol{R}^2 = R_{ik}^* S_{kl}^* S_{lm}^* R_{mn}^* R_{nj}^*$ etc. in Einstein notation.

The GEVH can be written with Reynolds stress as the subject of the equation by substituting Eq. (A1) into Eq. (2), which gives a sum of 11 terms instead. Coefficients $g^{(n)}$ must be tuned for the flow that Eq. (A1) is to be deployed on. A tensor basis neural network (TBNN) can be trained to predict these as demonstrated by Ling et al. [12]. The standard TBNN inputs for the invariant input layer to predict $g^{(n)}$ are the invariants of $S_{ij}^*$ and $R_{ij}^*$ given in Eq. (A2). After $g^{(n)}$ have been predicted, they must be elementwise multiplied with $\boldsymbol{T}^{(n)}$ to give $b_{ij}$ as shown in Eq. (A1). Therefore, tensors $\boldsymbol{T}^{(n)}$ must be supplied to the TBNN as a set of inputs. The standard TBNN tensor inputs are $\boldsymbol{T}^{(1)}$ to $\boldsymbol{T}^{(10)}$ given in Eq. (A1) and they are delivered to the tensor input layer.

## Appendix B: Hyperparameter Tuning

Hyperparameter tuning was performed via grid searches on the non-zonal TBNN and TKENN with combinations of the following discrete hyperparameter settings:

**Table 7** Hyperparameter Tuning Settings

| Hyperparameters | Discrete settings |
| --- | --- |
| No. of hidden layers | 2, 5, 10, 20 |
| No. of hidden nodes per hidden layer | 5, 10, 25, 50 |
| Activation functions | Rectified linear unit (ReLU), Exponential linear unit (ELU), Swish |
| Batch size | 16, 32, 64, 128, 256 |
| Learning rate (LR) scheduler factor (with LR = 0.01) | 0.98, 0.995 |

Three-fold cross-validation of the training and validation datasets was incorporated in these grid searches, as only a limited number of Re number cases were available for training and validation [52]. For every combination of hyperparameter values, three separate instances of the NN were trained and validated. In each instance, the validation dataset was chosen from one of Re = 1800, 4500 and 5400, and the NN was trained on the remaining two cases. The choice of validation dataset was rotated amongst the three available cases, such that a different case was used as the validation dataset in each NN instance. For a specific combination of hyperparameter values, three final validation errors were therefore produced, and these were averaged to give a mean final validation error. The combinations in **Table 8** and **Table 9** gave the lowest mean final validation errors, where the optimal hyperparameter set was ranked first:

**Table 8** Top Ranking TBNN Hyperparameter Combinations

| No. of hidden layers | No. of hidden nodes per hidden layer | Activation functions of hidden nodes | Batch size | Learning rate scheduler factor | Mean final validation error |
| --- | --- | --- | --- | --- | --- |
| 2 | 50 | Swish | 256 | 0.98 | $7.67 \times 10^{-2}$ |
| 5 | 10 | ELU | 128 | 0.98 | $7.72 \times 10^{-2}$ |
| 2 | 50 | Swish | 128 | 0.98 | $7.85 \times 10^{-2}$ |
| 2 | 10 | ReLU | 64 | 0.98 | $7.89 \times 10^{-2}$ |
| 2 | 50 | ReLU | 256 | 0.98 | $7.92 \times 10^{-2}$ |

**Table 9** Top Ranking TKENN Hyperparameter Combinations

| No. of hidden layers | No. of hidden nodes per hidden layer | Activation functions of hidden nodes | Batch size | Learning rate scheduler factor | Mean final validation error |
| --- | --- | --- | --- | --- | --- |
| 5 | 10 | ELU | 256 | 0.98 | 1.149 |
| 10 | 25 | ReLU | 32 | 0.98 | 1.158 |
| 2 | 25 | ReLU | 256 | 0.98 | 1.163 |
| 5 | 5 | ELU | 32 | 0.98 | 1.166 |
| 2 | 25 | Swish | 128 | 0.98 | 1.167 |